\documentclass[preprint,prd,aps,tighten,nofootinbib,amssymb]{revtex4}

\usepackage{graphicx}
\usepackage{epsfig,amssymb,amsmath}

\pdfoutput=1
\usepackage{subfigure}
\usepackage[colorlinks=true,linkcolor=blue,citecolor=blue]{hyperref}
\usepackage{float}

\usepackage{ulem}

\makeatletter
\def\Hline{%
\noalign{\ifnum0=`}\fi\hrule \@height 2pt \futurelet
\reserved@a\@xhline}
\makeatother

\newcommand{\beq}{\begin{equation}}
\newcommand{\eeq}{\end{equation}}
\newcommand{\bea}{\begin{eqnarray}}
\newcommand{\eea}{\end{eqnarray}}
\newcommand{\bear}{\begin{array}}
\newcommand {\eear}{\end{array}}
\newcommand{\bef}{\begin{figure}}
\newcommand {\eef}{\end{figure}}
\newcommand{\bec}{\begin{center}}
\newcommand {\eec}{\end{center}}

\begin{document}
\draft
\tighten
\preprint{CTPU-14-15, TU-988, IPMU14-0356}

\title{\large \bf
Degenerate spectrum in the neutrino mass anarchy with Wishart matrices and implications for $0\nu\beta\beta$ and $\delta_{\rm CP}$
}
\author{
	Kwang Sik Jeong\,$^{a,b}$\footnote{email:ksjeong@pusan.ac.kr},
	Naoya Kitajima\,$^{c}$\footnote{email:kitajima@tuhep.phys.tohoku.ac.jp},
	Fuminobu Takahashi\,$^{c,d}$\footnote{email:fumi@tuhep.phys.tohoku.ac.jp}
}
\affiliation{
	$^a$ Department of Physics, Pusan National University, Busan, 609-735, Korea\\
	$^b$ Center for Theoretical Physics of the Universe, IBS, Daejeon 305-811, Korea\\
	$^c$ Department of Physics, Tohoku University, Sendai 980-8578, Japan\\
	$^d$ Kavli IPMU, TODIAS, University of Tokyo, Kashiwa 277-8583, Japan
}

\vspace{2cm}

\begin{abstract}
We show that a degenerate neutrino mass spectrum can be realized in the neutrino mass
anarchy hypothesis, if the neutrino Yukawa and right-handed neutrino mass matrices 
are given by the Wishart matrix, i.e. products of $N \times 3$ rectangular random matrices, 
whose eigenvalue distribution 
tends to degenerate for large $N$.
The mixing angle and CP phase distributions 
are determined  by either the Haar measure of U(3) or that of SO(3). 
We study how large $N$ is allowed to be without tension with the observed neutrino mass squared differences,
and find that the predicted value of $m_{ee}$ can be within the reach of future $0\nu\beta\beta$ experiments
especially for $N$ on the high side of the allowed range. 
\end{abstract}

\maketitle

\section{Introduction}
\label{sec:intro}
The standard model (SM) of particle physics has been overwhelmingly successful for decades,
and the long-sought Higgs boson, the last missing piece of the SM, was finally 
discovered at the LHC~\cite{Aad:2012tfa,Chatrchyan:2012ufa}. 
Despite the great success of the SM, there are many puzzles left unanswered; one of them
is the origin of the flavor structure. 

While neutrinos are massless in the SM, atmospheric and solar neutrino oscillation experiments revealed that neutrinos have 
tiny but non-zero masses (see e.g. Refs.~\cite{Forero:2014bxa,Gonzalez-Garcia:2014bfa} for the latest results). 
In particular,  a mild mass hierarchy and large mixing angles for the neutrino sector 
are in sharp contrast with  quarks and charged leptons.
If we are to understand the neutrino flavor structure
based on symmetry principles, it seems to require
rather contrived flavor models.\footnote{
While it is possible to understand the hierarchical mass pattern of quarks and charged
leptons based  on symmetry principles,  a variety of flavor symmetries and charge assignments are allowed.
For an alternative approach without flavor symmetry, see e.g. Ref.~\cite{Hall:2007zj}.
}
The observed large mixing angles 
rather suggest structureless mass matrix for neutrinos, implying that 
all the neutrino species have the same quantum number.

The squared mass differences and mixing angles are measured by various neutrino oscillation experiments 
\cite{An:2012eh,Abe:2012tg,Ahn:2012nd,Abe:2011sj,Abe:2011fz,Adamson:2011qu}
and the recent best-fit values for normal (inverted) hierarchy are \cite{Forero:2014bxa}
\beq
\label{obs}
	\begin{split}
		&\Delta m^2_{21} = 7.60 \times 10^{-5}~{\rm eV}^2,~~|\Delta m^2_{31}| = 2.48~(2.38) \times 10^{-3}~{\rm eV}^2 \\[1mm]
		&\sin^2\theta_{12} = 0.323,~~\sin^2\theta_{23} = 0.567~(0.573),~~\sin^2\theta_{13}= 0.0234~(0.0240),
	\end{split}
\eeq
and the favored value of the Dirac CP phase is around $3 \pi/2$.
Besides the neutrino oscillation experiments, 
further information can be obtained from the cosmic microwave background (CMB) observations and the neutrinoless 
double beta decay ($0\nu \beta\beta$) experiments. In particular, 
the CMB observations by Planck, WMAP and other ground-based experiments 
set the upper limit on the sum of the neutrino masses as  $\sum m_i < 0.66~{\rm eV}~(95\%{\rm\,CL})$~\cite{Ade:2013zuv}.

One of the attractive explanations for the observed large neutrino mixing 
is the neutrino mass anarchy~\cite{Hall:1999sn,Haba:2000be,deGouvea:2003xe,deGouvea:2012ac},
which gained momentum especially after the discovery of a non-zero value of $\theta_{13}$
by the Daya-Bay experiment~\cite{An:2012eh}. The basic idea of the neutrino mass anarchy is simple.
Suppose that all the Yukawa couplings and/or right-handed neutrino masses are determined by a UV theory,
which has  a sufficiently large landscape of vacua. If  each coupling is allowed to take  values of order unity
in the landscape,  the Yukawa couplings and/or  right-handed neutrino masses may be modeled by 
some functions of random matrices. 
Note that, as emphasized in Ref.~\cite{Haba:2000be},  the neutrino mass anarchy tells us nothing 
about the weighting functions, and therefore, one has to choose an appropriate one to evaluate
the probability distribution of the neutrino masses.
The simplest and the most studied form is the linear measure:
\beq
h,~M \; \sim \; X,
\label{linear}
\eeq
where the neutrino Yukawa matrix $h$ as well as  the right-handed neutrino mass matrix $M$ are proportional 
to $3 \times 3$ random matrices represented by $X$. 
Phenomenological and cosmological aspects of the neutrino mass anarchy
have been studied; e.g.  two of the present authors (KSJ and FT) studied
the implications of neutrino mass anarchy for leptogenesis   in
 Ref.~\cite{Jeong:2012zj}, and it was also recently revisited in Ref.~\cite{Lu:2014cla}.
 See also Refs.~\cite{Feldstein:2011ck,Heeck:2012fw,Bai:2012zn} for phenomenological study of the neutrino mass anarchy
 with a various number of right-handed neutrinos.

In the neutrino mass anarchy hypothesis, the mixing angle and CP phase distributions are determined by the invariant 
Haar measure of the underlying symmetry group such as U(3) or SO(3)~\cite{Haba:2000be}, 
 independently of the adopted weighting function, and so, there are rather robust predictions.
Interestingly, the observed large mixing angles can be nicely explained in the neutrino mass anarchy~\cite{deGouvea:2012ac}.\footnote{
See, however, Refs.~\cite{Altarelli:2012ia,Bergstrom:2014owa} and references therein.}
On the other hand,  the neutrino mass spectrum depends sensitively on the weighting functions.
In the case of the linear measure, 
normal mass hierarchy is highly favored over the inverted or quasi-degenerate one.
In addition,  the observed mild hierarchy of the mass squared differences can be nicely explained by the
neutrino mass anarchy together with the seesaw 
mechanism~\cite{Hall:1999sn,Haba:2000be}.  The estimated $m_{ee}$
turned out to be too small to be detected by future $0\nu\beta\beta$ experiments~\cite{Jeong:2012zj}, 
but this result can be modified for more general measure functions~\cite{Jenkins:2008ms}.\footnote{
Our analysis is different from Ref.~\cite{Jenkins:2008ms} in which the adopted measure is not applicable to the 
case of the seesaw mechanism with neutrino mass anarchy. 
 }

In this letter we study the next simplest possibility: the neutrino Yukawa couplings and the right-handed neutrino
masses are given by the random matrix squared, or more precisely, the Wishart matrices:
\beq
h,~M\;\sim \; X^\dag X {\rm~or~}X^T X
\eeq
where $X$ represents $N\times 3$ complex or real random matrices. In general, $N$ does not have to be equal to $3$.
For $N > 3$, the neutrino Yukawa and right-handed neutrino mass matrices are given by
products of rectangular matrices. 
We shall see that the observed neutrino mass squared differences can be explained for $N \lesssim 35$. 
Interestingly, the eigenvalue distribution of the Wishart matrix tends to be degenerate  for large $N$.\footnote{
A similar behavior can be seen in the singular value distributions of the $n \times 3$ neutrino Yukawa matrix if one introduces $n~(> 3)$ right-handed neutrinos \cite{Heeck:2012fw}.
However, the eigenvalues of the $n\times n$ Majorana mass matrix obeying the linear measure are 
more repulsive than in the case of the Wishart matrix.
Thus, while the resultant neutrino masses are also degenerate to some extent for large $n$, the degeneracy is weaker than in the case of the Wishart matrix.}
Therefore,
quasi-degenerate neutrino mass spectrum can be realized in the neutrino mass anarchy with
the Wishart matrix if $N \gg 3$, which should be contrasted to the case of the linear measure (\ref{linear}).
We will discuss its implications for the $0\nu\beta\beta$ experiments.
We will also show that the mixing angle and CP phase distributions of
our scenario are determined by either the Haar measure of U(3) or that of SO(3).

The rest of this letter is organized as follows. In Sec.~\ref{sec:anarchy} we first explain our set-up and
see how the neutrino mass spectrum changes as the size of the rectangular matrices $N$ increases. 
Then we study the implication for the Dirac CP phase and 
the $0\nu\beta\beta$ experiments. The last section is devoted for discussion and conclusions.

\section{Neutrino mass anarchy}
\label{sec:anarchy}
In this section, we consider the neutrino mass anarchy based on the Wishart matrices
as a simple extension of the linear measure.
We focus on the case of the Majorana neutrino mass with the seesaw 
mechanism~\cite{Minkowski:1977sc,Yanagida:1979as,Ramond:1979py,Glashow:1979nm}.\footnote{
Our set-up can be straightforwardly applied to the case of the Dirac neutrino mass,
and most of our results (except for the $0\nu\beta\beta$) will remain 
qualitatively valid. In particular, the quasi-degenerate spectrum can be realized.}

\subsection{Preliminaries}
\label{subsec:set-up}

The seesaw Lagrangian is given by 
\beq
	\mathcal{L} = f_{ij} \bar{e}_{Ri} \ell_j \tilde{H} + h_{ij} \bar{N}_i \ell_j H +\frac{1}{2} M_{ij} \bar{N}_i \bar{N}_j + {\rm h.c.},
\eeq
where $\ell$, $H(\tilde{H})$, $e_R$ and $N$ are respectively the left-handed lepton doublet, the Higgs doublet (its SU(2) 
conjugate),  the right-handed charged leptons and the right-handed neutrinos, 
$f_{ij}$, $h_{ij}$ are Yukawa matrices for charged leptons and neutrinos respectively 
and $M_{ij}$ represents the Majorana mass matrix for right-handed neutrinos. The subscripts represent the generation,
$i,j = 1,2,3$.

Let us first diagonalize the charged lepton Yukawa matrix as
\bea
 f &= 
U_{fR}^\dag D_e U_{fL}
\eea
with
\bea
D_e&\equiv 
\left(
\bear{ccc}
y_e & 0& 0\\
0&y_\mu&0\\
0&0&y_\tau
\eear
\right),
\eea
where $U_{fR}$ and $U_{fL}$ are  unitary matrices, and $y_{e,\mu,\tau} (>0)$ denote the charged lepton Yukawa couplings.\footnote{
Throughout this letter we do not try to interpret the charged lepton mass hierarchy in our scheme because there could be additional 
selection (anthropic) effects.
} 
In the basis where the charged lepton Yukawa matrix is diagonal, the Lagrangian becomes
\beq
	\mathcal{L} = (y_{\alpha} \delta_{\alpha \beta}) \bar{e}_{R\alpha} \ell_\beta \tilde{H} 
	+ h_{i\alpha} \bar{N}_i \ell_\alpha H +\frac{1}{2} M_{ij} \bar{N}_i \bar{N}_j + {\rm h.c.},
\eeq
where $\alpha, \beta$ run over the lepton flavor indices $(e,\mu,\tau)$, and we have defined
\bea
\label{h}
h_{i\alpha} &\equiv h_{ij} \left(U_{fL}^\dag\right)_{j \alpha}, \\
\ell_\alpha &\equiv \left(U_{fL}\right)_{\alpha i} \ell_i,\\
e_{R\alpha} &\equiv \left(U_{fR}\right)_{\alpha i} e_{Ri}.
\eea

After the Higgs field acquires the vacuum expectation value (VEV), one obtains
the effective Lagrangian for active neutrinos  by integrating out the heavy right-handed neutrinos, 
\beq
	\mathcal{L}_{\rm eff}=  - \frac{1}{2} (m_\nu)_{\alpha \beta} \nu_\alpha \nu_\beta + {\rm h.c.},
\eeq
where $\nu_\alpha$ are the light left-handed neutrinos, and the neutrino mass matrix is given by 
\beq
\left(m_\nu \right)_{\alpha \beta} = v^2 \left(h^T\right)_{\alpha i} \left(M^{-1}\right)_{ij} h_{j \beta}
\eeq
with $v \simeq 174~{\rm GeV}$ being the VEV of the Higgs field.
The neutrino mass matrix  $m_\nu$ is generically a complex-valued symmetric matrix, and 
it can be diagonalized by a unitary matrix $U_{\rm MNS}$ as
\beq
	m_\nu = U_{\rm MNS}^*
	\begin{pmatrix} m_1 & 0 & 0 \\ 0 & m_2 & 0 \\ 0 & 0 & m_3 \end{pmatrix} U_{\rm MNS}^\dag.
\eeq
Here $m_1$, $m_2$ and $m_3$ are real and positive values with $m_1 < m_2 < m_3$.
This numbering is for the normal hierarchy, whereas in the inverted hierarchy case,
one should relabel them as  $m_3 \to m_2$, $m_2 \to m_1$ and $m_1 \to m_3$ 
in order to compare our results with the observations (\ref{obs}). 
In fact, however, mostly either normal or quasi-degenerate (normal-ordering) mass hierarchy is 
 realized in our scheme, and so, the inverted hierarchy case is practically negligible. 

The neutrino oscillation experiments provide us with only the squared mass differences, $\Delta m^2_{ij} = m^2_i-m^2_j$.
In order to compare our results with observations,  we use the dimensionless parameter $R$ defined by the ratio of the 
squared mass difference between the heaviest and the second
heaviest neutrinos to that between the second heaviest and the lightest ones:
\beq
	R = \frac{\Delta m^2_{21}}{\Delta m^2_{32}}~({\rm normal})~~\text{or}~~ \frac{\Delta m^2_{13}}{\Delta m^2_{21}} ~({\rm inverted}).
\eeq
The observed value of $R$  is given by
$R \sim 1/30$ for normal-ordering hierarchy and $R \sim 30$ for inverted hierarchy.

The neutrino mixing matrix $U_{\rm MNS}$ can be expressed in terms of the mixing angles, $\theta_{ij}$, with
$(i,j)=(1,2),(2,3)$ and $(3,1)$, 
and the Dirac and Majorana CP phases, $\delta$, $\alpha_{21}$ and $\alpha_{31}$ 
after absorbing the unphysical phases by redefinition of the fields, and it is conventionally written as
\beq
	U_{\rm MNS} = 
	\begin{pmatrix}
		c_{12} c_{13} & s_{12} c_{13} & s_{13} e^{-i\delta} \\[1mm]
		-s_{12} c_{23}-c_{12} s_{23} s_{13} e^{i\delta} & c_{12} c_{23} - s_{12} s_{23} s_{13} e^{i\delta} & s_{23} c_{13} \\[1mm]
		s_{12} s_{23} - c_{12} c_{23} s_{13} e^{i\delta} & -c_{12} s_{23} - s_{12} c_{23} s_{13} e^{i\delta} & c_{23} c_{13}
	\end{pmatrix}
	\times {\rm diag} \big(1,~e^{i\frac{\alpha_{21}}{2}},~e^{i\frac{\alpha_{31}}{2}} \big) 
\eeq
where we abbreviate $\sin\theta_{ij}$ and $\cos\theta_{ij}$ as $s_{ij}$ and $c_{ij}$, respectively, and
the mixing angles and the CP phases satisfy $\theta_{ij} \in [0, \pi/2)$ and $\delta, \alpha_{21}, \alpha_{31}, \in [0, 2\pi)$.

\subsection{Neutrino mass anarchy based on  the Wishart matrices}
\label{subsec:wishart}

In the neutrino mass anarchy hypothesis with the linear measure,
both $h_{i\alpha}$ and $M_{ij}$ are taken to be proportional to $3 \times 3$ complex(or real)-valued random 
matrices (cf. Eq.~(\ref{linear})).  The unitary matrix $U_{fL}$ does not affect the probability distributions
of the mixing angles and the CP phases, as they are fixed by the Haar measure of U(3) (SO(3)). 
This is the simplest possibility, but it remains unknown how  the randomness for these matrices is originated in the
landscape.  In fact,  there are various other basis-independent choices for these matrices. 
Here we consider the next-to-simplest set-up, in which the neutrino Yukawa matrix and Mayorana mass matrix 
consist of products of random matrices:\footnote{
If the neutrino Yukawa couplings and the right-handed neutrino masses are given by 
$h \sim F^T F$ and $M \propto G^T G$, where $F$ and $G$ are complex-valued $N \times 3$ random matrices of order unity,
there is no degeneracy in the eigenvalues. We do not pursue this case in this letter.
}

\beq
	h_{ij} = \frac{y_\nu}{N} \left(F^\dagger F\right)_{ij},~~~ M_{ij} = \frac{M_0}{2N} \left(G^\dagger G + G^T G^*\right)_{ij},
	\label{h_and_M}
\eeq
where $F$ and $G$ are $N \times 3$ complex (or real) random matrices of order unity, and $y_\nu$ and
$M_0$ represent the typical neutrino Yukawa couplings and the right-handed neutrino masses.
For $y_\nu = {\cal O}(1)$, $M_0 \sim 10^{15}~{\rm GeV}$ is suggested by the neutrino oscillation experiments and
the seesaw mechanism. 
Note that the above form of the neutrino Yukawa couplings is given in the original basis, and one has to multiply it with 
 the unitary matrix $U_{fL}$ in the basis where the charged lepton Yukawa matrix is diagonal (see Eq.~(\ref{h})). 
 This however does not affect the final mixing and CP phase distributions just as in the  previous case.\footnote{
In general, any Yukawa matrix can be written as a product of a Hermitian matrix and a unitary matrix by the
polar decomposition theorem. Here we consider a case where the Hermitian matrix is of the Wishart-type random 
matrix.
 } 

The above form of $h_{ij}$ and $M_{ij}$ imply that they are given by the so-called Wishart matrix.
Specifically, we will take $F$ and $G$ as a chiral Gaussian Unitary (Orthogonal) Ensemble, i.e. the Gaussian measure, where each element
follows a complex(real)-valued Gaussian distribution with zero mean and a variance of unity. In this case, the basis-independence is automatically assured \cite{Lu:2014cla}.
The measure for the eigenvalues ($\lambda_i$) of the complex and real Wishart matrix composed of $N \times 3$ random matrices are respectively known as
\beq
	\prod^{3}_{i>j} |\lambda_i - \lambda_j|^2 \prod^{3}_{i=1} \lambda_i^{N-3} d\lambda_i~~\text{(complex)},~ \prod^{3}_{i>j} |\lambda_i - \lambda_j| \prod^{3}_{i=1} \lambda_i^{(N-4)/2} d\lambda_i ~~\text{(real)}.
\eeq
The first factor $|\lambda_i - \lambda_j|$ represents the repulsive nature,  and this effect is (partially)
canceled by the second factor $\lambda_i^{N-3}$ or $\lambda_i^{(N-4)/2}$.
For large $N$, the eigenvalues of $h$ and $M$ tend to be highly degenerate due to the second
factor proportional to $\lambda_i^{N-3}$ or $\lambda_i^{(N-4)/2}$.\footnote{
Instead of the Gaussian measure, one can adopt an arbitrary basis independent measure for the Wishart matrix. For instance one may multiply $({\rm tr}[G^\dag G])^p$ with
the measure. In this case, the eigenvalue distribution are modified, but the eigenvalues remain to be
degenerate for large $N$ as long as the measure contains positive powers of the eigenvalues.
}
As a result,  the light neutrino masses are also expected to be degenerate, which is difficult to realize in the case of the 
linear measure.  As we shall see, however, $N$ cannot be arbitrarily large because the predicted value of $R$ tends
to be too large compared to the observed value, $R\sim 1/30$, for large $N$.

\subsection{Mass spectrum, mixing angles and CP phases}
\label{subsec:mass}

\begin{figure}[t]
\centering
\subfigure[]{
\includegraphics [width = 7.5cm, clip]{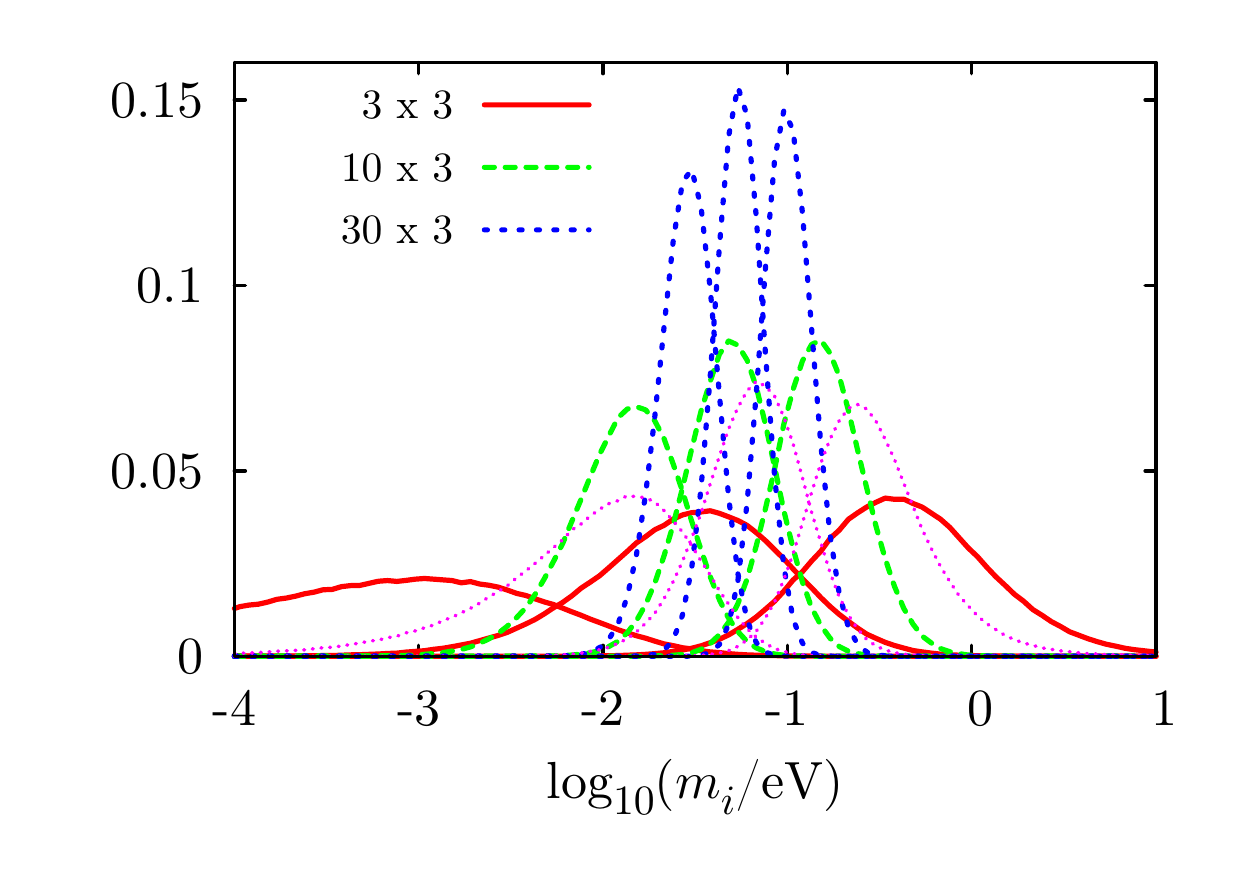}
\label{subfig:mass}
}
\subfigure[]{
\includegraphics [width = 7.5cm, clip]{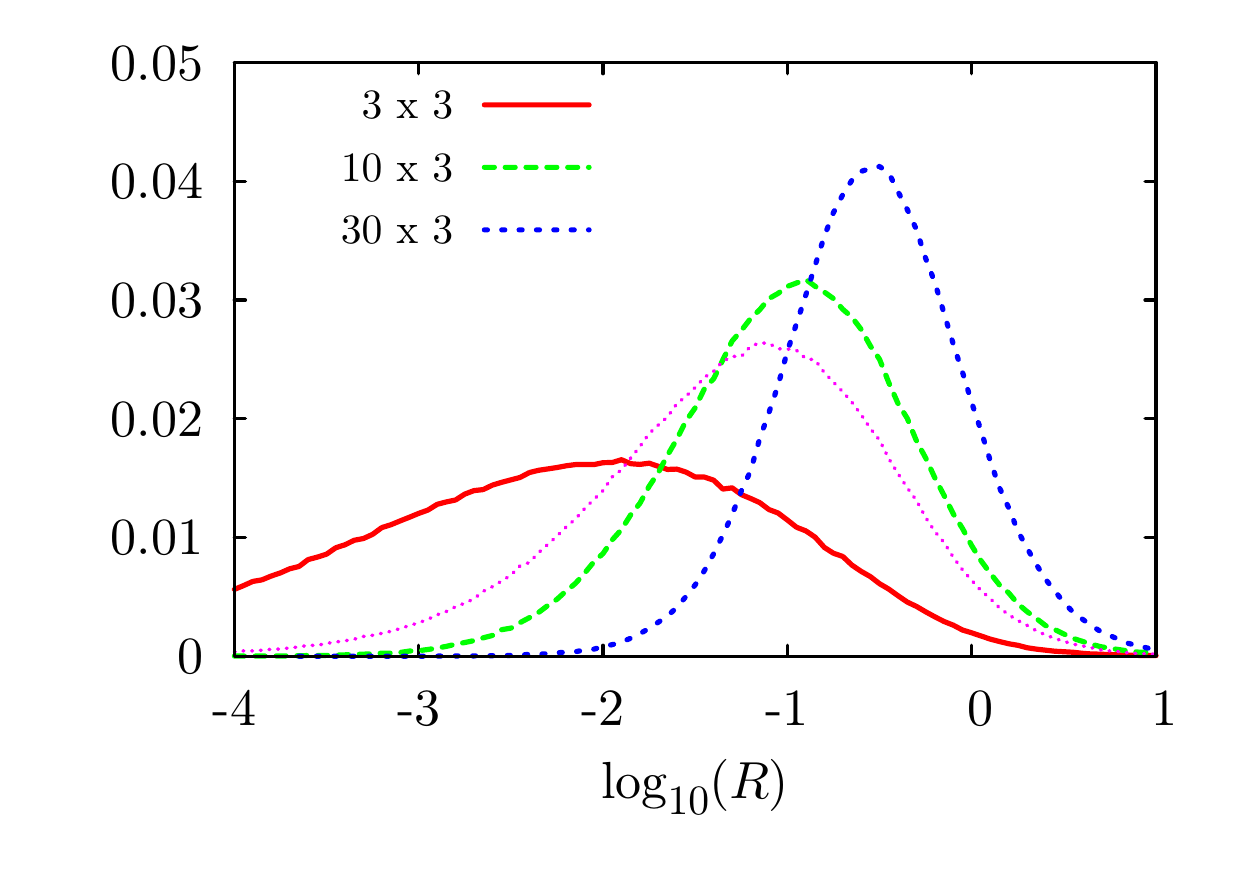}
\label{subfig:R}
}
\caption{
	Probability distributions of the neutrino masses (left) and $R$ (right) for complex Wishart matrices are shown.
	The solid red, dashed green and dotted blue lines correspond to the case with $N =$ 3, 10 and 30 respectively, 
	while the magenta lines represent the  anarchy with the linear measure.
	Here we have taken $y_\nu=1$ and $M_0 = 10^{15}~{\rm GeV}$.
}
\label{fig:dist}
\end{figure}

\begin{figure}[t]
\centering
\subfigure[]{
\includegraphics [width = 7.5cm, clip]{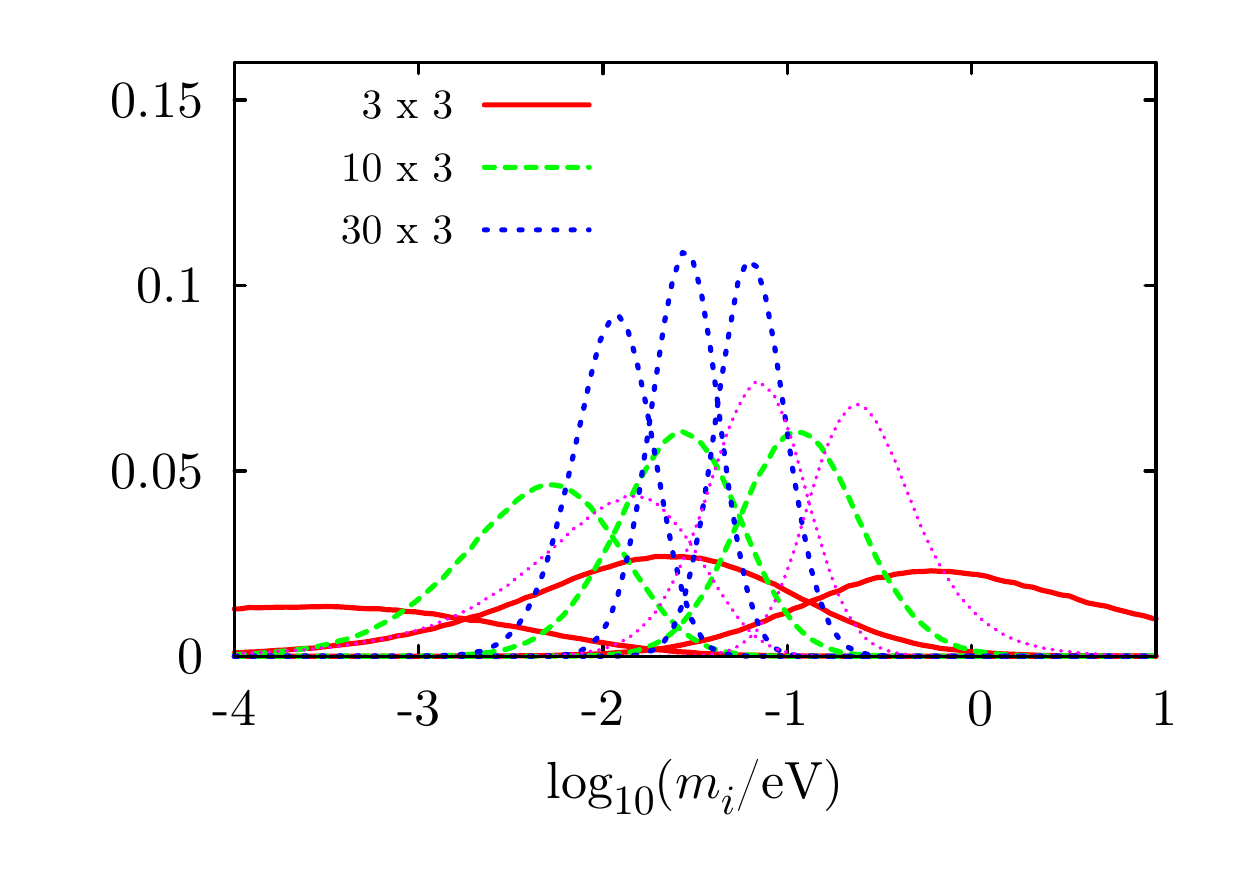}
\label{subfig:mass_real}
}
\subfigure[]{
\includegraphics [width = 7.5cm, clip]{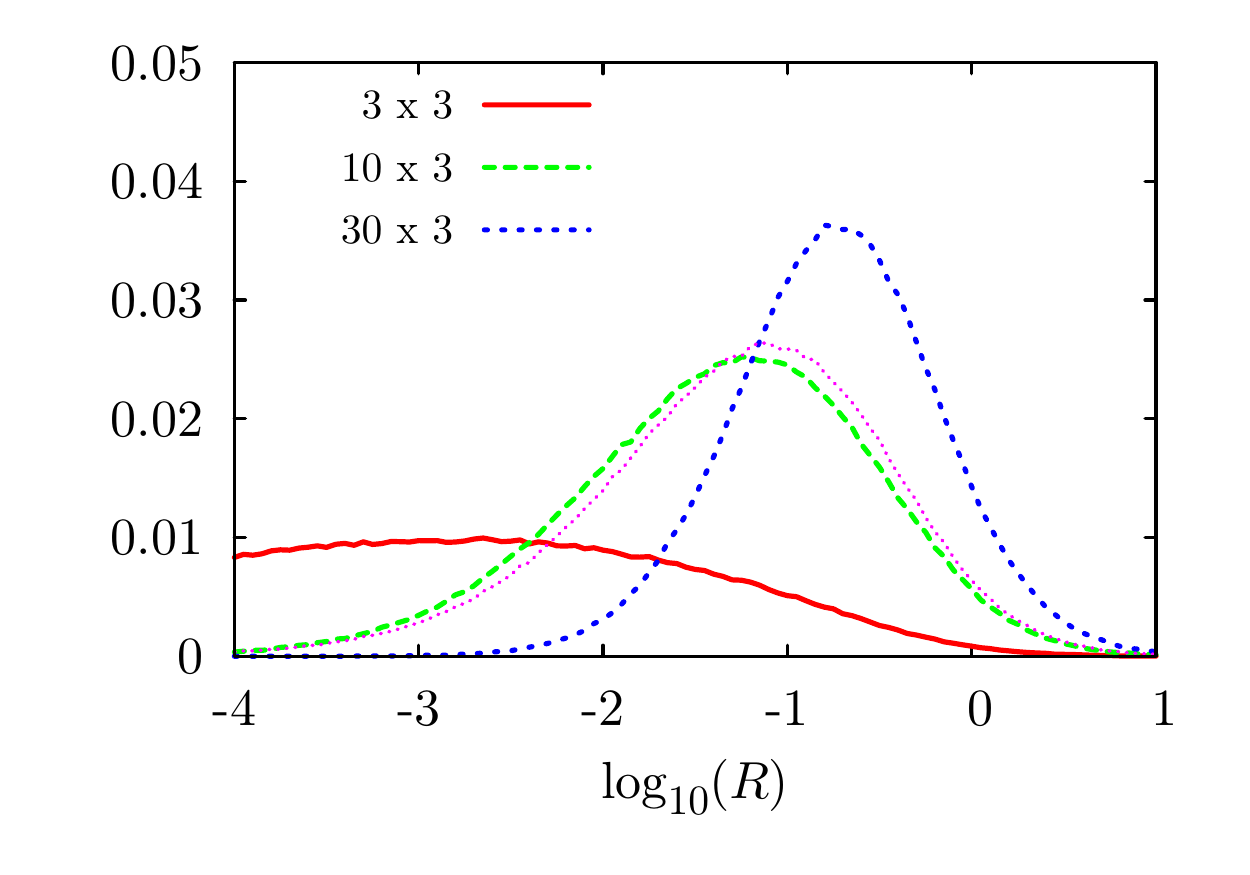}
\label{subfig:R_real}
}
\caption{
Same as Fig.~\ref{fig:dist} but for real Wishart matrices. 
}
\label{fig:dist_real}
\end{figure}

We have performed numerical calculations of the neutrino mass anarchy based on the Wishart matrices.
Specifically, we have generated  $10^6$ $N \times 3$ complex (real) random matrices, $F$ and $G$,  
to obtain the distributions of neutrino masses, mixing angles and CP violating phases. The results
 are shown in Figs.~\ref{fig:dist} and \ref{fig:dist_real} corresponding to the complex and real Wishart matrices, respectively.
We have varied $N$ as $N = 3$ (solid red), $N = 10$ (dashed green), $N = 30$ (dotted blue), and we have set $y_\nu = 1$ and
  $M_0 = 10^{15}~{\rm GeV}$. Note that the distribution of $R$ in the right panel is independent of the choice of $y_\nu$ and $M_0$. 
For comparison, we show the results of the neutrino anarchy with the linear measure as the small-dotted magenta lines in each figure.
One can see that the neutrino mass distribution (Figs.~\ref{subfig:mass} and \ref{subfig:mass_real}) tends to be more degenerate as $N$ increases.
The probability distribution of $R$ is suppressed at $R > 1$, implying that the inverted hierarchy ($R \sim 30$) is
highly disfavored. Thus,  the neutrino mass hierarchy is either normal or quasi-degenerate (normal-ordering) in the  anarchy
based on the Wishart matrices.

\begin{figure}[th]
\centering
\subfigure[]{
\includegraphics [width = 7.5cm, clip]{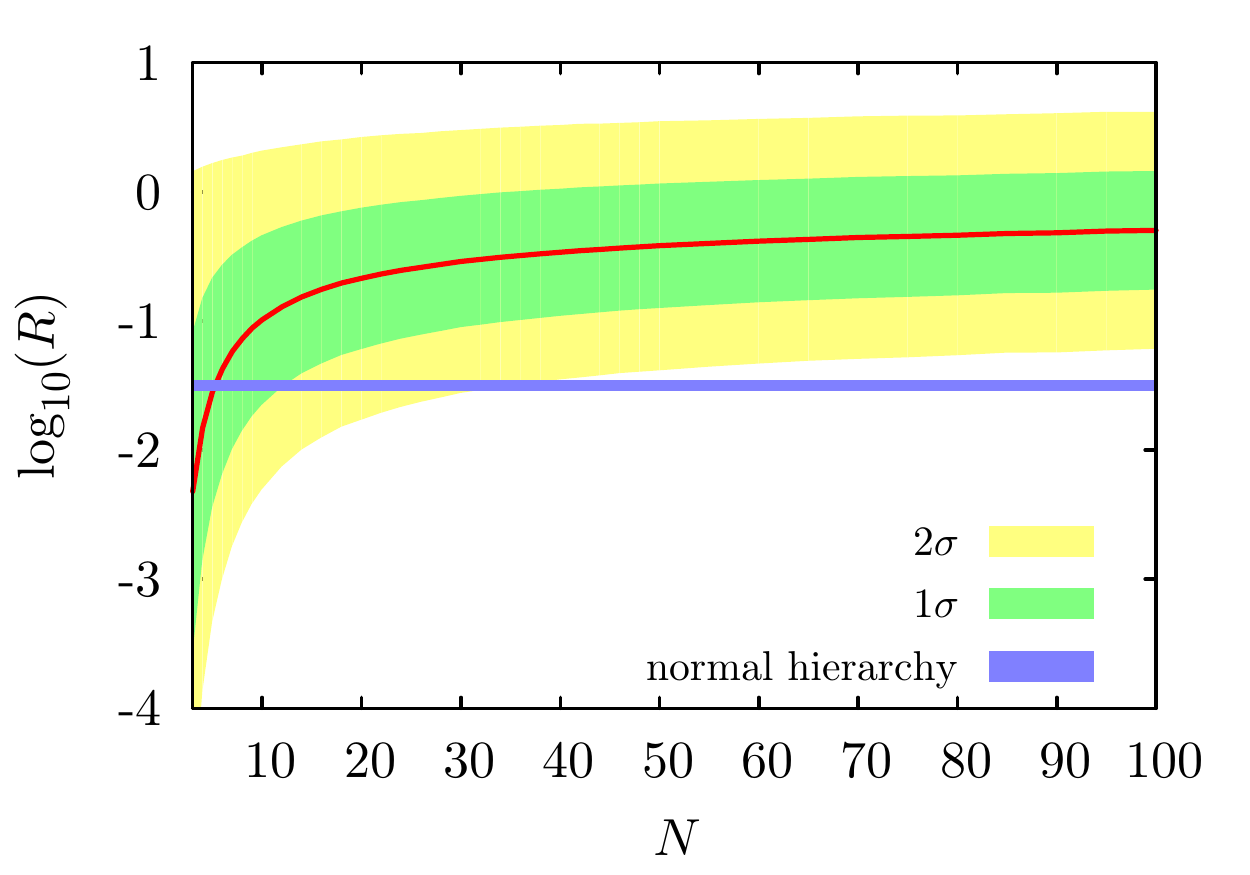}
\label{subfig:Rmean}
}
\subfigure[]{
\includegraphics [width = 7.5cm, clip]{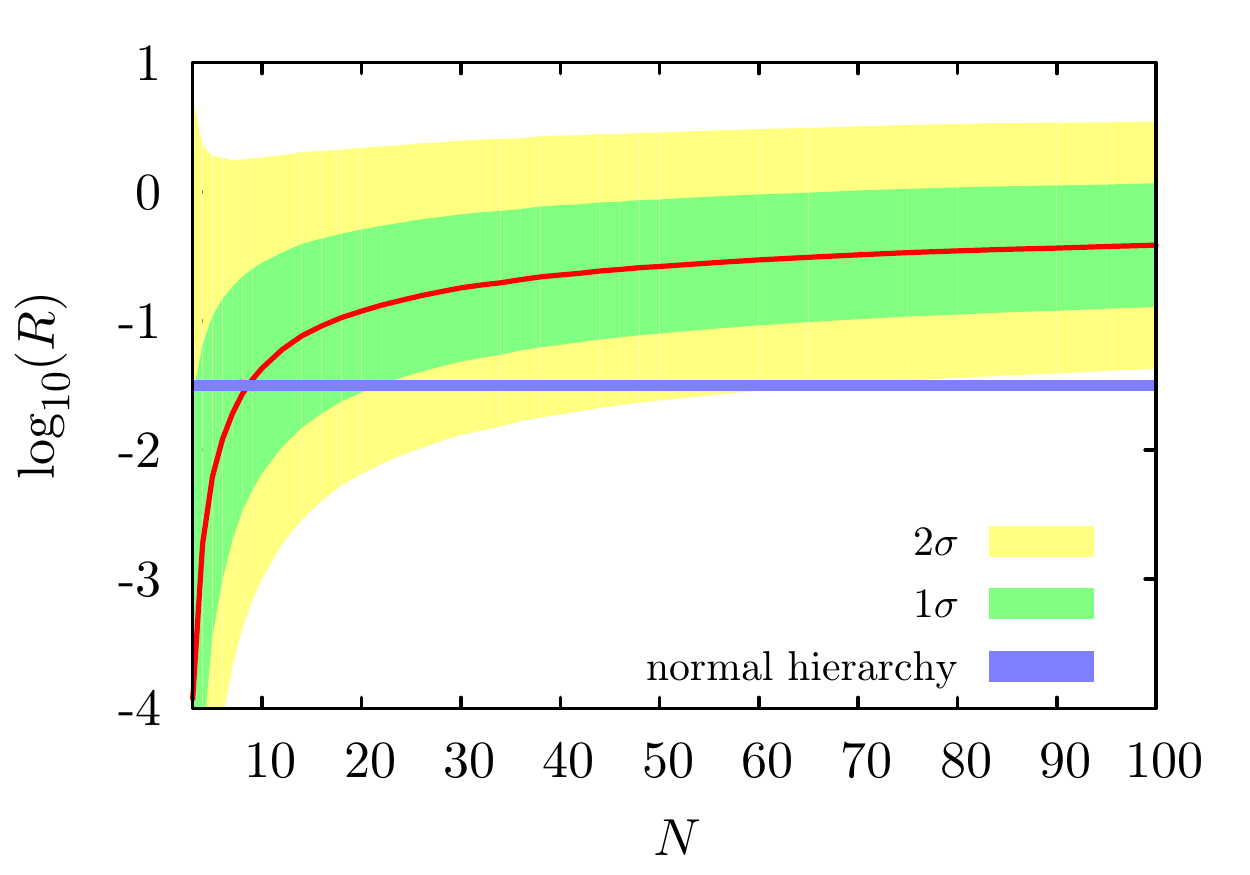}
\label{subfig:Rmean_real}
}
\caption{
	The mean value of $R$ with 1$\sigma$ (green region) and 2$\sigma$ (yellow region) error  as a function of $N$ 
	corresponding to complex (left) and real (right) random matrices.
	The blue-shaded region represents the experimental value with 2$\sigma$ uncertainty for the normal hierarchy.
}
\label{fig:Rmean}
\end{figure}

 Fig.~\ref{fig:Rmean} shows the mean value of $R$ as a function of $N$ with $1$ and  $2\sigma$ error bands.
It shows that the normal hierarchy ($R \sim 1/30$) is preferred over the inverted hierarchy ($R\sim 30$) and $N$ is bounded from above as $N \lesssim 35$ 
($N \lesssim 70$ for real Wishart matrices) in order to be consistent with the observations. 
This implies that, even if one considers the Wishart matrices, there is an upper bound on the degeneracy of the neutrino masses.
We will discuss its implications for the $0\nu\beta\beta$ experiments in the next subsection. 

\begin{figure}[th]
\centering
\subfigure[]{
\includegraphics [width = 7.5cm, clip]{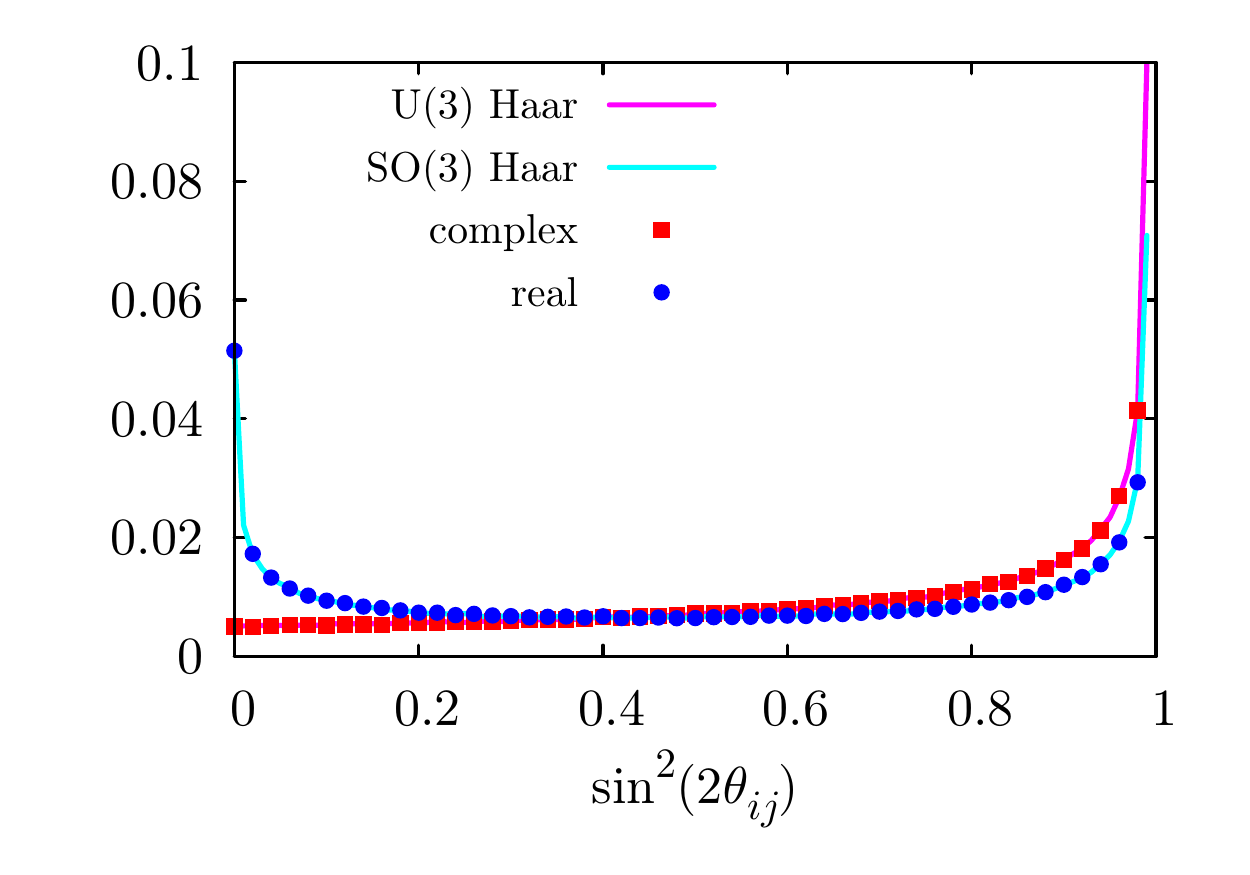}
\label{subfig:theta_ij}
}
\subfigure[]{
\includegraphics [width = 7.5cm, clip]{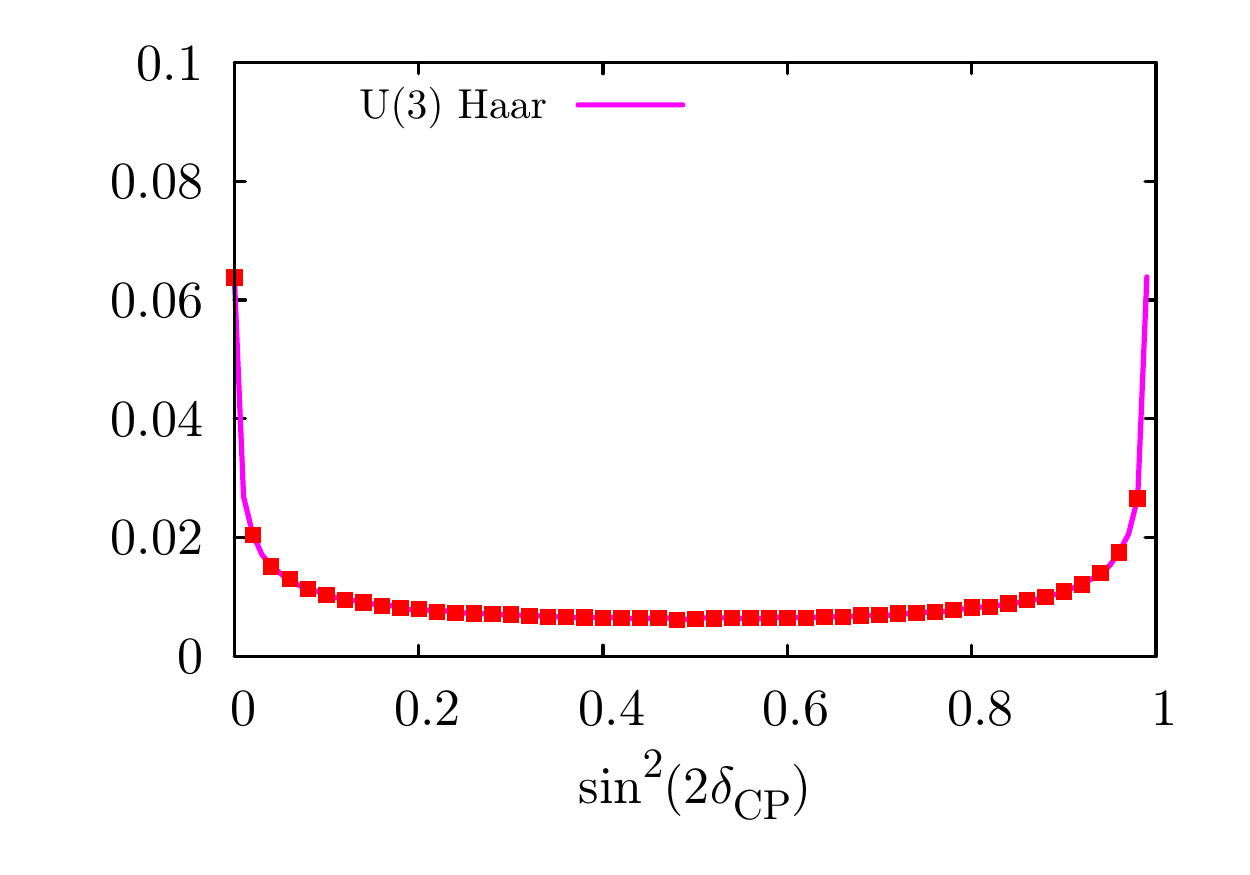}
\label{subfig:deltaCP}
}
\caption{
	Probability distributions of mixing angles (left) and CP violating phases (right).
	$\theta_{ij}$ represents $\theta_{12}$, $\theta_{23}$ and $\theta_{13}$ and $\delta_{\rm CP}$ represents $\delta$, $\alpha_{21}$ and $\alpha_{31}$.
	The red squares and blue circles correspond to complex and real Wishart matrices, respectively, 
	and the magenta and cyan lines correspond to the U(3) and SO(3) Haar measure respectively.
	We have taken $N = 30$, but the distributions are the same for a different value of $N$.
}
\label{fig:theta_delta}
\end{figure}

We can also see from Fig.~\ref{fig:theta_delta} that the mixing angle and CP phase distributions are determined by the Haar measure of U(3).
If the random matrices $F$ as well as the charged lepton Yukawa matrix are taken to be real, the resultant distribution is 
given by the Haar measure of SO(3). (The right-handed neutrino mass matrix is real by construction.)
In this case the Majorana CP phases vanish, and the Dirac CP phase $\delta$ takes a value of either
$0$ or $\pi$. 
We note that the currently favored value of  $\delta$ is about $3 \pi/2$ according to 
Ref.~\cite{Gonzalez-Garcia:2014bfa}, which corresponds to $\sin^2 2\delta =0$. Interestingly, the U(3) Haar measure results
in the probability distribution of $\delta$ peaked at $\sin^2 2\delta =0$.

\subsection{Neutrinoless double beta decay}
\label{subsec:0nu}
The Majorana nature of the neutrinos can be probed by the $0\nu \beta\beta$
experiments, which is sensitive to $m_{ee}$ defined by
\beq
\begin{split}
m_{ee} &\equiv  \left|\sum_{i=1}^{3}(U_{\rm MNS})^2_{ei}\,m_i \right| \\[1mm]
&= \bigg| m_1 (c_{12}c_{13})^2 + m_2 (s_{12}c_{13})^2 e^{i \alpha_{21}} + m_3 s_{13}^2 e^{i(\alpha_{31}-2 \delta)} \bigg|.
\end{split}
\label{eq:m_ee}
\eeq
The current upper bound on  $m_{ee}$ by the GERDA
experiment  using $^{76}$Ge  reads~\cite{Agostini:2013mzu}
\beq
m_{ee} \;\lesssim\;(0.2-0.4)\,{\rm eV}~~(90\%{\rm CL}).
\eeq
A similar bound was obtained by EXO-200 using $^{136}{\rm Xe}$~\cite{Albert:2014awa}, and
a slightly better bound has been recently obtained by the 
KamLAND-Zen experiment as~\cite{TheKamLAND-Zen:2014lma}
\beq
m_{ee}\; \lesssim\; (0.14-0.28)~{\rm eV}~~(90\%{\rm CL}).
\eeq
The next-generation experiment is expected to
reach the level of  $m_{ee} \simeq 0.01$~eV \cite{Dell'Oro:2014yca}.

\begin{figure}[t]
\centering
\subfigure[$ N = 10$]{
\includegraphics [width = 7.5cm, clip]{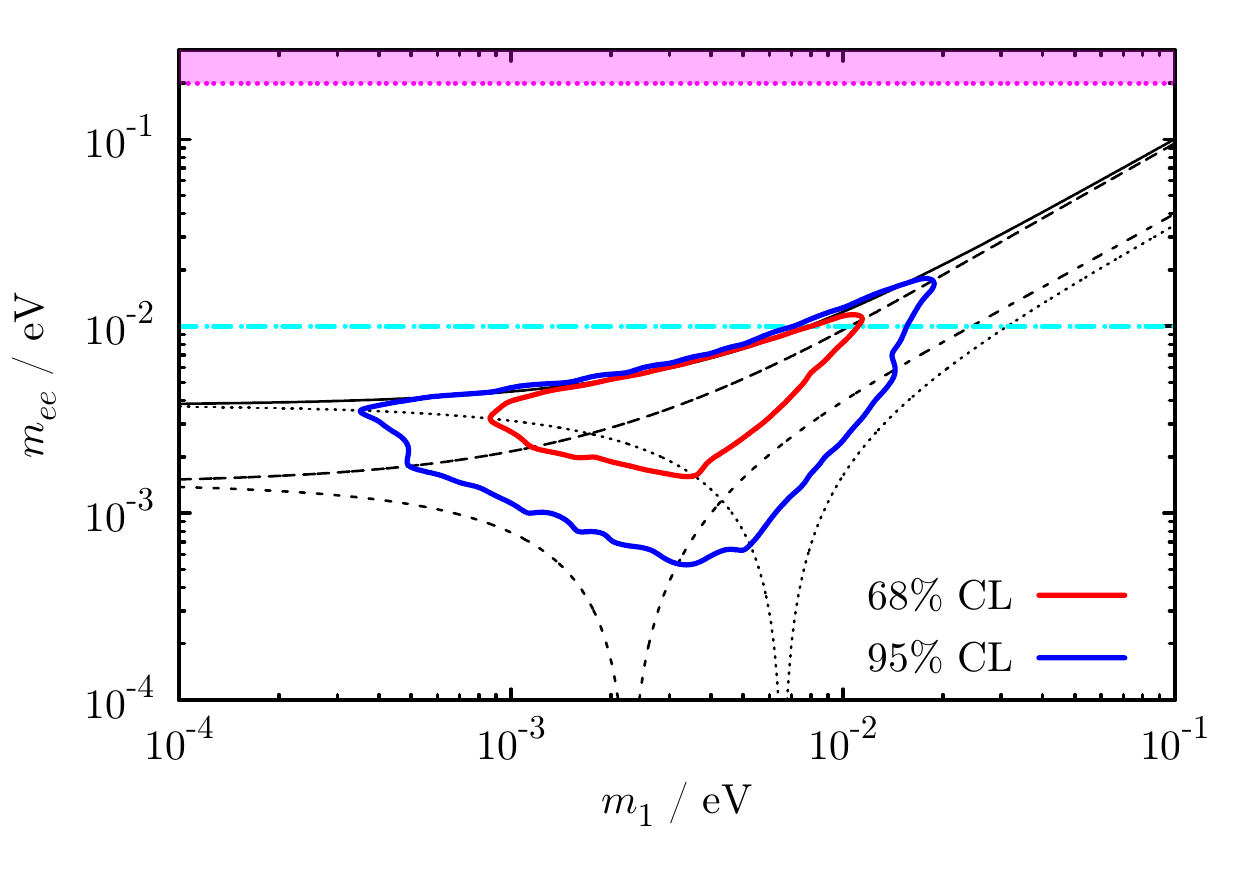}
\label{subfig:m_ee10}
}
\subfigure[$ N = 30$]{
\includegraphics [width = 7.5cm, clip]{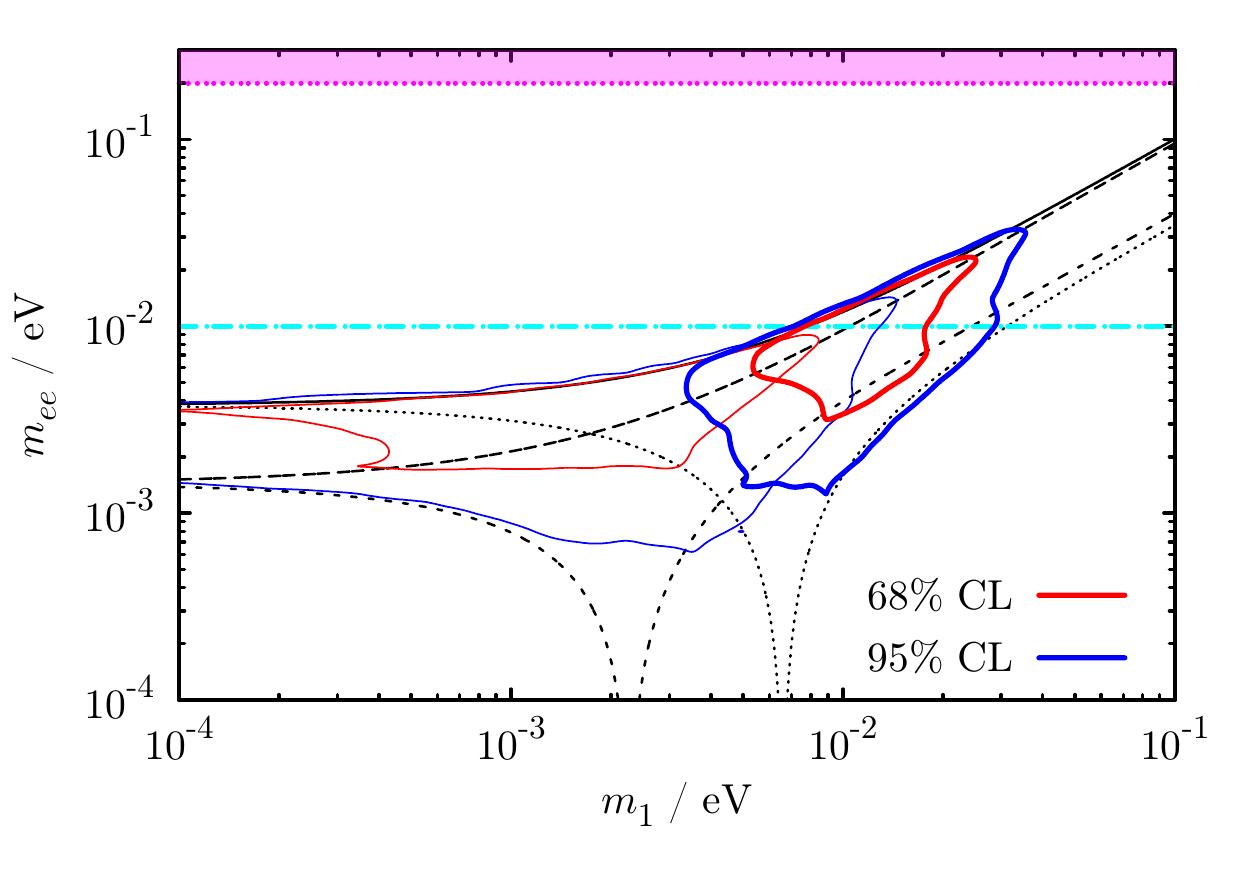}
\label{subfig:m_ee30}
}
\caption{
	Contours of probability distribution on $m_{ee}$--$m_1$ plane for $N = 10$ (left) and $N = 30$ (right), where 
	the mixing angles are set to be the best-fit values.
	The red and blue contours correspond to 68\% and 95\% CL respectively, and for comparison, the case of the linear measure
	is shown by the thin red and blue contours in the right panel.
	The black  curves with various line types correspond to the normal hierarchy for
	best fit values of the  neutrino mass differences and mixing angles with vanishing CP phases; 
	 $(e^{i \alpha_{21}}, e^{i(\alpha_{31}-2 \delta)}) = (+1,+1)$, $(+1,-1)$, $(-1,+1)$ and $(-1,-1)$ from
	 top to bottom at $m_1 \gtrsim 10^{-2}$\,eV.
	The horizontal dashed (cyan) line represents the sensitivity of future experiment, while the shaded (magenta) region is excluded
	by the current experiments.
}
\label{fig:m_ee}
\end{figure}

\begin{figure}[t]
\centering
\subfigure[$N = 10$]{
\includegraphics [width = 7.5cm, clip]{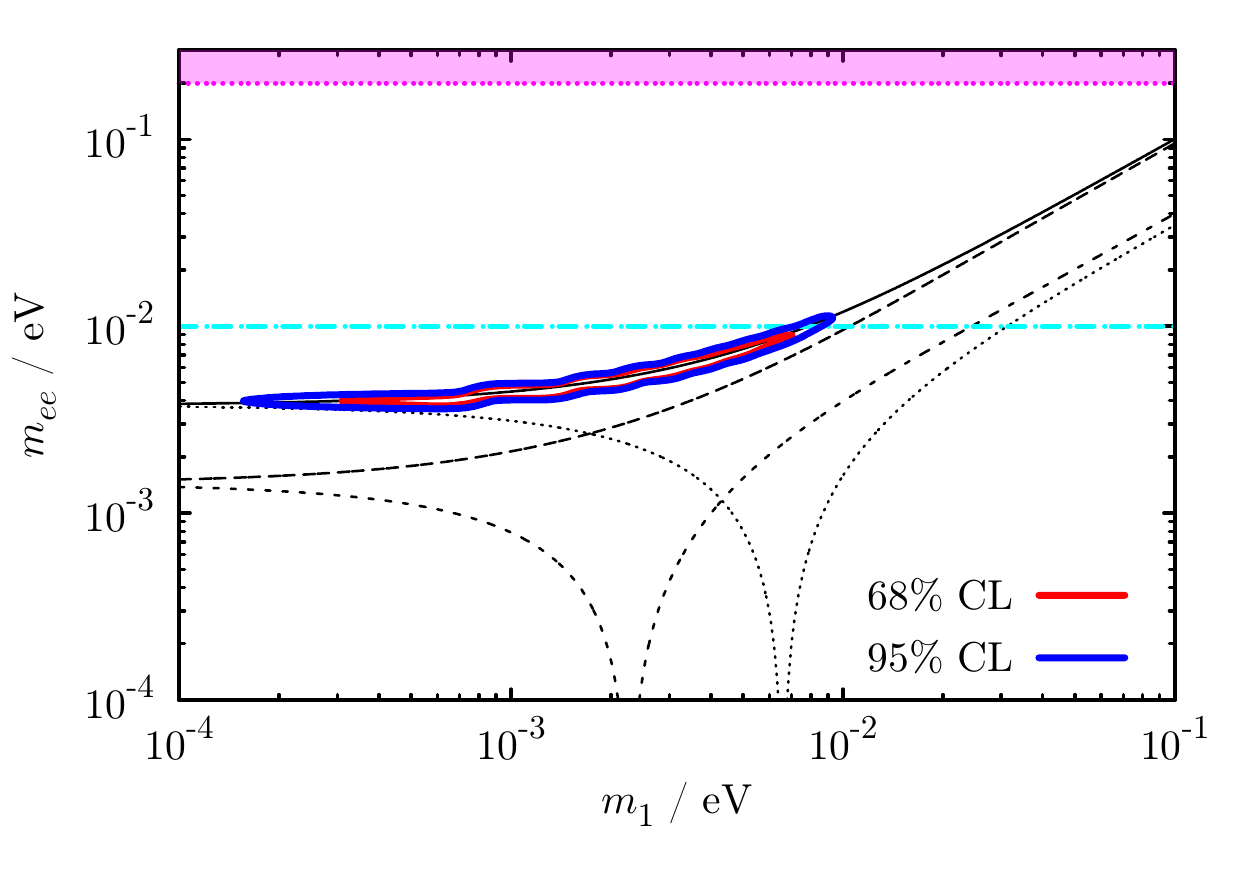}
\label{subfig:m_ee_real10}
}
\subfigure[$N = 30$]{
\includegraphics [width = 7.5cm, clip]{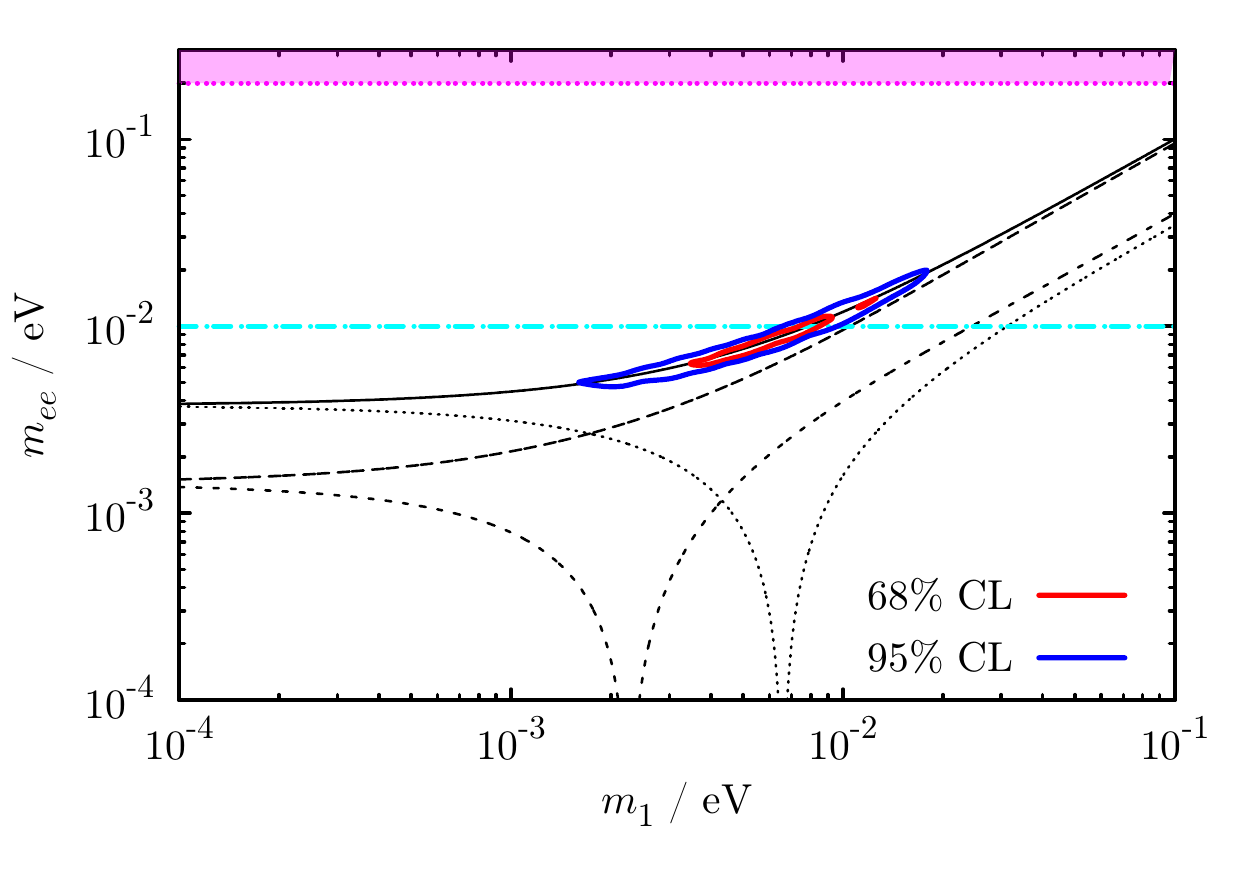}
\label{subfig:m_ee_real30}
}
\caption{
Same as Fig.~\ref{fig:m_ee} but for real Wishart matrices. Here we have chosen
the case of $(e^{i \alpha_{21}}, e^{i(\alpha_{31}-2 \delta)}) = (+1,+1)$.
}
\label{fig:m_ee_real}
\end{figure}

\begin{figure}[t]
\centering
\subfigure[]{
\includegraphics [width = 7.5cm, clip]{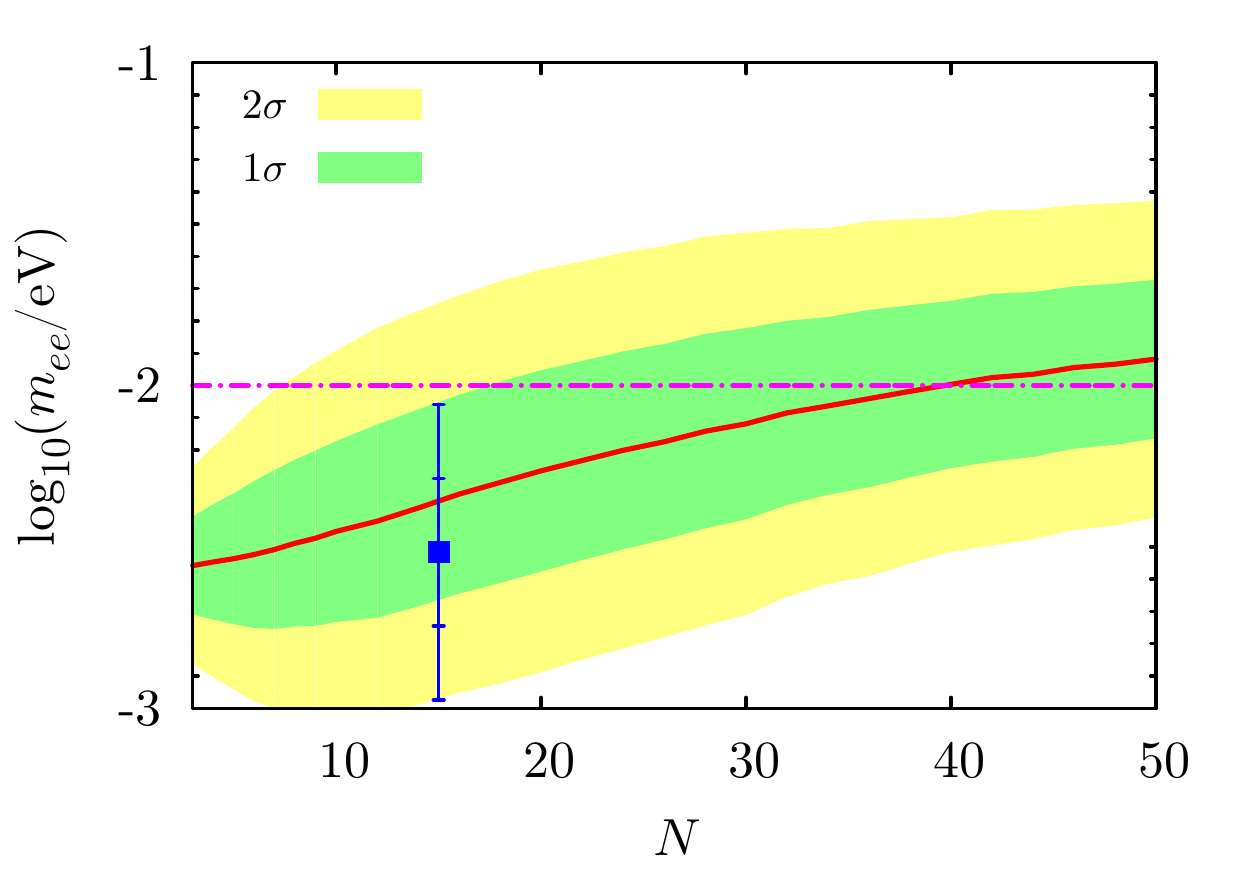}
\label{subfig:m_ee_mean}
}
\subfigure[]{
\includegraphics [width = 7.5cm, clip]{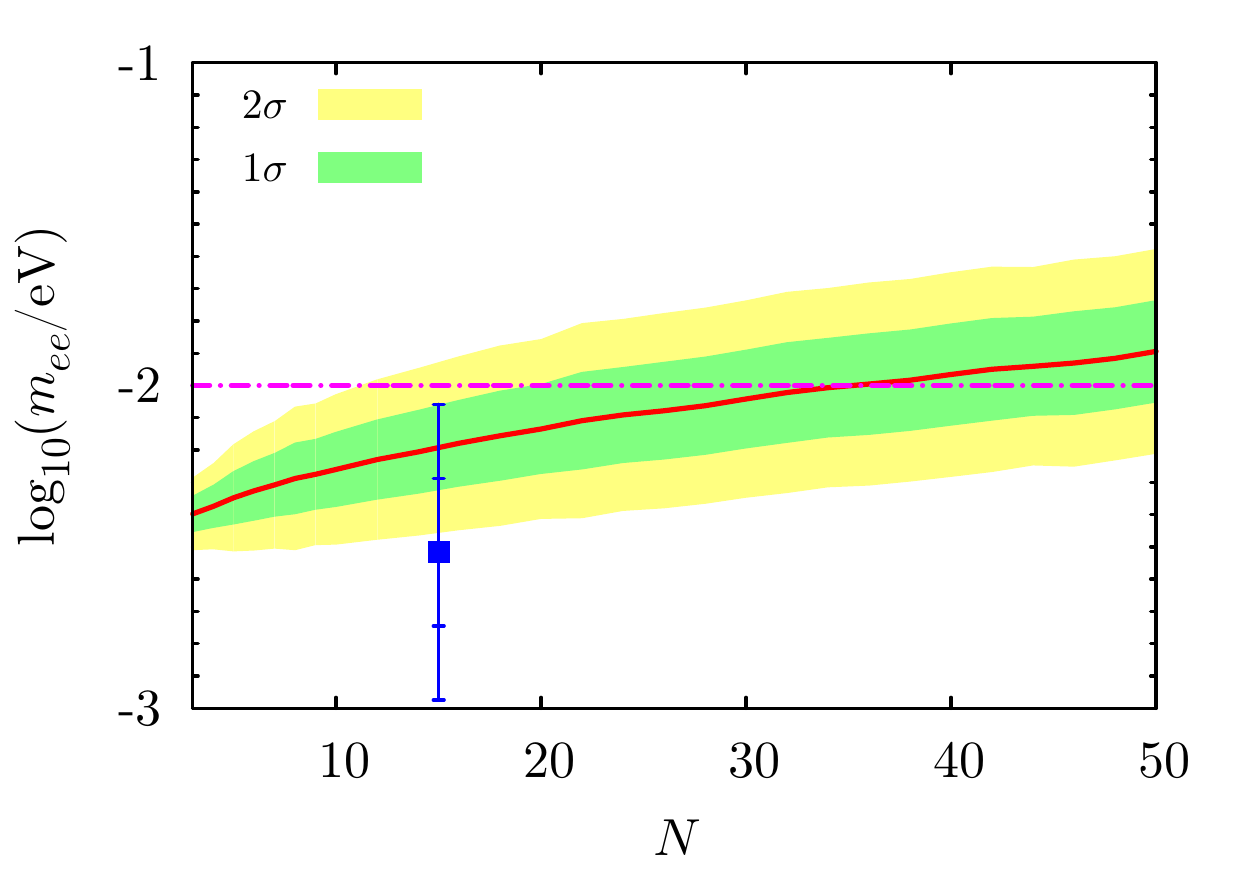}
\label{subfig:m_ee_mean_real}
}
\caption{
The mean value of $m_{ee}$ (red solid line) with 1$\sigma$ (green) and 2$\sigma$ (yellow) uncertainties as a function of $N$, for complex (left) and real (right) random matrices. 
The horizontal dashed (magenta) line represents the sensitivity of future experiment.
The blue point with an error bar represents the one for the linear measure with 1$\sigma$ and 2$\sigma$ uncertainties. (The position in the horizontal axis is arbitrary.)
}
\label{fig:m_ee_mean}
\end{figure}

We show the predicted range of $m_{ee}$ in the $m_{ee}$--$m_1$ plane in Fig.~\ref{fig:m_ee} (complex Wishart) and Fig.~\ref{fig:m_ee_real} (real Wishart),
where we have taken $N = 10$ and $30$. 
We have generated $10^7$ Wishart matrices 
and extracted the subset satisfying the observed $R$ (within 2$\sigma$) 
and $M_0$ is adjusted to realize the best fit value of $\Delta m_{21}^2$. The mixing angles are also adjusted to the best fit values.
The thick red (blue) lines are contours of equal probability in which 68\% (95\%) of the data points are contained. 
For comparison, we similarly show the prediction of the linear measure case as thin red (blue) lines in the right panel of Fig.~\ref{fig:m_ee}.
The black lines with various line types represent  $m_{ee}$ for best-fit values of the neutrino mass differences and mixing angles 
with vanishing CP-violating phases: $(e^{i \alpha_{21}}, e^{i(\alpha_{31}-2 \delta)}) = (+1,+1)$, $(+1,-1)$, $(-1,+1)$ and $(-1,-1)$ 
from	 top to bottom at $m_1 \gtrsim 10^{-2}$\,eV. The horizontal dashed (cyan) line represent the sensitivity of the future experiment, while the shaded (magenta) region is excluded by the current experiments.
We also show the statistical mean value of $\log_{10}(m_{ee}/\mathrm{eV})$ with 1 and 2$\sigma$ uncertainties as a function of $N$ in Fig.~\ref{fig:m_ee_mean}.
Since a quasi-degenerate mass spectrum is more likely for large values of $N$, 
relatively large $m_{ee} (\gtrsim 0.01\,{\rm eV})$ is realized with a greater probability compared to the case of the linear measure 
and a larger fraction of the parameter space will be accessible by the near future experiments. Note however that, since $N$ is
bounded from above in order to be consistent with observations, there is an upper bound on the neutrino mass degeneracy. As a
result, $m_{ee}$ cannot be arbitrarily large even in the case with the Wishart matrices (i.e., $m_{ee} \lesssim $ a few tens meV).

\section{Discussion and Conclusions}
\label{sec:conc}

In this letter we have studied in detail the neutrino mass anarchy hypothesis with the Wishart matrices,
where the neutrino Yukawa matrices and right-handed neutrino masses
are given by products of $N\times 3$ random rectangular matrices. The mixing angle and CP phase distributions
are determined  by the Haar measure of U(3) or SO(3), depending on whether the Wishart matrices are
complex or real. Interestingly, for $N \gg 3$, the eigenvalues of the Wishart matrix tend to be confined in a narrow range. 
As a result, compared to the case of the neutrino mass anarchy
with the linear measure, the neutrino mass spectrum becomes more compressed, in particular,
a quasi-degenerate (normal-ordering) neutrino mass spectrum can be easily realized without resort to introducing additional
constraints (such as successful leptogenesis~\cite{Jeong:2012zj,Lu:2014cla}) or an ad hoc choice of the
weighting function. We have studied how large $N$ is allowed to be in order to give a reasonable fit to the observed neutrino
mass squared differences and found that $N$ is allowed to be as large as $35$ for complex Wishart matrices and $70$ for real
Wishart matrices.
We have also studied implications of our scenario for the $0\nu\beta\beta$ experiment, and shown that the predicted $m_{ee}$  can be within the reach of the future experiments with a larger probability than the case of the linear measure, 
especially if $N$ is on the high side of the allowed range.

Let us  discuss if we can understand the structure of the couplings based on symmetry principles.
First let us regard the random matrices $F$ and $G$ as moduli fields whose VEVs can take various values
determined by a UV theory. To be specific we assume that all the couplings are real, and impose O(N)$\times$O(3)
flavor symmetry, under which the ordinary leptons and right-handed neutrinos transform as ${\bf 1} \times {\bf 3}$
while $F$ and $G$ transform as ${\bf N}\times {\bf 3}$. The lepton doublets and the right-handed neutrinos are assumed to transform as ${\bf 3}$ under O(3).
Then, the following combination
\beq
F^T F, ~~~G^T G
\eeq
are $3 \times 3$ matrices, which  transform as bifundamental under $O(3)$.
Once each component of $F$ and $G$ develops a non-zero VEV, the above matrices
give rise to the neutrino Yukawa couplings and the Majorana masses. If the UV theory is sufficiently complicated,
the VEVs of $F$ and $G$ may be modeled by random matrices. 
Thus, the above combination $F^T F$ and $G^T G$ play the same role of the simple random matrix 
in the case of the linear measure.  One can see that how much the above set-up is more complicated than in the case of the linear measure.

In principle one can add an unit matrix  to the Yukawa and the right-handed
neutrino matrices, satisfying the flavor symmetries. If the contribution of the unit matrix is negligible  compared
to that of $F$ and $G$, our results in the text approximately remain unchanged in this case. On the other hand,
if the unit matrix contribution becomes significant, the mass eigenvalues become more degenerate, whereas the 
mixing angle distribution is still determined by the SO(3) Haar measure.\footnote{This argument suggests another extension
of the neutrino mass anarchy with the linear measure: one may add a unit matrix (with a numerical coefficient) to the
neutrino Yukawa and the right-handed neutrino mass matrices, leading to degenerate mass spectra while the
mixing angle and CP phase distribution are still given by the U(3) or SO(3) Haar measure. }

We would like to emphasize here that the above argument explains only the structure of the interactions,
not the reason why the measure is proportional to the random matrix squared. The essence of the neutrino mass
anarchy hypothesis is the (statistical) equivalence between different neutrino flavors, and it tells us nothing about the weighting measure functions. The simplest and most studied function is the linear measure, but, there is 
no compelling reason to choose this measure other than simplicity. In general, the weighting measure could be
some complicated function of the random matrices. In this sense, our choice of the measure is the
next simplest possibility.

So far we have focused on the neutrino mixing, mass, and CP phase distributions in the neutrino mass
anarchy with the Wishart matrices.  It will be interesting to study cosmological aspects of our scenario, 
especially in context with leptogenesis, as an extension
of the analysis of Ref.~\cite{Jeong:2012zj}. In particular,  in contrast to the case of the linear measure,
the right-handed neutrinos tend to be degenerate in mass, leading to resonant leptogenesis~\cite{Pilaftsis:1997dr}.
The typical mass difference scales as $(M_2-M_1)/(M_2+M_1) \sim 1/\sqrt{N}$, and so, we expect that 
an enhancement of the lepton asymmetry by a factor of $5$ or so for $N=30$. If  the value of $N$ is different
between the neutrino Yukawa and right-handed neutrino mass matrices, this factor may be even more enhanced.
We however expect that it is hard to realize the enhancement by many orders of magnitude in our scenario
because the eigenvalues still repel each other even in the limit of large $N$. This difficulty may be eased by
allowing a contribution proportional to the unit matrix. 
We leave the detailed analysis of leptogenesis in this case for future work. 

As pointed out in Refs.~\cite{Hall:1999sn,Haba:2000be}, one can impose a flavor symmetry without modifying the
predictions for the light neutrino masses: for instance we can introduce a flavor symmetry on the right-handed neutrinos.
Then, while the right-handed neutrinos are hierarchical due to the non-trivial flavor charges, the light neutrinos remain
degenerate.

We can consider a possibility that the neutrino Yukawa and the right-handed neutrino mass matrices are given by
 a more complicated function(s) of random matrices, such as the Wishart matrices squared, and so on. Alternatively
 one may consider sparse random matrices. It may be interesting to study these possibilities and their implications
 for the neutrino masses and CP phases. 
 
\section*{Acknowledgment}
This work was supported by  JSPS Grant-in-Aid for
Young Scientists (B) (No.24740135 [FT]), 
Scientific Research (A) (No.26247042 [FT]), Scientific Research (B) (No.26287039 [FT]), 
 the Grant-in-Aid for Scientific Research on Innovative Areas (No.23104008 [NK, FT]),  and
Inoue Foundation for Science [FT].  This work was also
supported by World Premier International Center Initiative (WPI Program), MEXT, Japan [FT].
KSJ was supported by IBS under the project code, IBS-R018-D1.

\bibliographystyle{/Users/kitajimanaoya/bib/bst/utphys}
\bibliography{/Users/kitajimanaoya/bib/bib}

\end{document}